# Enabling Continuous THz Band Coverage via Precise Electron Beam Tailoring in Free-electron Lasers

## Author Information

Yin Kang[1,2,3,4], Tong Li[5], Zhen Wang[1], Yue Wang[4], Cheng Yu[1], Weiyi Yin[1], Zhangfeng Gao[1], Hanghua Xu[1], Hang Luo[1], Xiaofan Wang[6], Jian Chen[1], Taihe Lan[1], Xiaoqing Liu[1], Jinguo Wang[1], Huan Zhao[1], Fei Gao[1], Liping Sun[1], YanYan Zhu[1], Yongmei Wen[1], Qili Tian[5], Chenye Xu[7], Xingtao Wang[1], Jiaqiang Xu[1], Zheng Qi[1], Tao Liu[1], Bin Li[1,3,7], Lixin Yan[5], Kaiqing Zhang[1,3*], Chao Feng[1,3*], Bo Liu[1,3] & Zhentang Zhao[1,3*]

## Affiliations

[1]Shanghai Advanced Research Institute, Chinese Academy of Sciences, Shanghai, China

[2]Shanghai Institute of Applied Physics, Chinese Academy of Sciences, Shanghai, China

[3]University of Chinese Academy of Sciences, Beijing, China

[4]Zhangjiang Laboratory, Shanghai, China

[5]Department of Engineering Physics, Tsinghua University, Beijing, China

[6]Institute of Advanced Science Facilities, Shenzhen, China

[7]School of Physical Science and Technology, ShanghaiTech University, Shanghai, China

Correspondence to: Chao Feng, fengc@sari.ac.cn; Kaiqing Zhang, zhangkq@sari.ac.cn; Zhentang Zhao, zhaozt@sari.ac.cn

## Abstract

High-power, continuously tunable narrowband terahertz (THz) sources are essential for advancing nonlinear optics, THz-driven material dynamics, and ultrafast spectroscopy. Conventional techniques typically impose a trade-off between pulse energy and frequency tunability. Here, we introduce a novel free-electron laser approach that overcomes these limitations by pre-modulating a relativistic electron beam with a frequency-beating laser pulse and leveraging bunch compression along with collective effects to enhance microbunching. Experimental results demonstrate that this technique generates narrowband THz emission with continuous frequency tunability from 7.8 to 30.8 THz, achieving pulse energies up to 385 $\mu J$ while maintaining spectral bandwidths between 7.7% and 14.7%. Moreover, the method exhibits exceptional robustness and scalability, highlighting its unique ability to bridge the long-standing THz gap and offering a promising solution for diverse cutting-edge scientific applications.

## Introduction

Narrowband terahertz (THz) radiation has emerged as an essential tool for probing low-energy quantum phenomena, offering direct access to fundamental material excitations such as molecules rotations and resonances, lattice vibrations, and superconducting oscillations [1-3]. In recent years, high-power THz radiation has garnered increasing interest for driving strong-field excitations, including THz-induced quantum material control, THz-driven magnetization dynamics, nonlinear THz optics and spectroscopy, THz-triggered chemistry reactions and single-shot THz bioimaging [4-9]. Although several high-power THz sources with pulse energies reaching the millijoule ($mJ$) level have been reported [10-14], their operational frequencies are typically fixed. On the other hand, numerous narrowband THz sources have been demonstrated [15-17], but they are generally limited to pulse energies on the order of microjoules ($\mu J$) level and exhibit restricted frequency tunability.

Free-electron lasers (FELs), which couple relativistic electron beams and electromagnetic fields, can produce tunable, narrowband radiation across a broad frequency range from X-rays to THz [18]. Conventional THz FELs typically rely on optical cavity oscillators amplified by sequential, elongated low-current electron bunches, thereby limiting their energy conversion efficiency and peak power [19-21]. Similarly, single-pass THz FELs based on self-amplified spontaneous emission suffer from low efficiency due to radiation diffraction and slippage effect, where the electron beam lags behind the electromagnetic wave in the undulator [22]. To enhance energy conversion efficiency, a single-pass THz FEL operating under zero-slippage resonance has been developed, offering the frequency tunability from 0.4 to 1 THz [23, 24].

Narrowband THz radiation in the 0.1-10 THz range can be generated via superradiant emission from ultrashort electron bunches traversing an undulator, provided the electron beam is compressed to lengths significantly shorter than the corresponding radiation wavelength [25-27]. However, due to the slippage effect, the progressive lag between the electron bunch and the emitted radiation, superradiant emission cannot be continuously amplified, thereby limiting the peak field strength. Moreover, compressing a high-charge electron beam to durations much shorter than 100 fs (required for frequencies above 10 THz) while preserving beam qualities remains a challenge for low-energy accelerators. To achieve both higher field strength and higher-frequency THz superradiation, the generation of ultrashort electron bunch trains becomes essential [28-32]. To our knowledge, no narrowband, high-intensity THz source has yet been demonstrated that continuously covers the 5-30 THz gap.

In this work, we introduce a high-power, narrowband THz FEL mechanism driven by a precisely tailored electron beam generated through optical frequency beating. By exploiting collective effects in the accelerator, our approach produces electron bunch trains with programmable spacing via longitudinal phase-space manipulation at relativistic energies, thereby mitigating the adverse effects of longitudinal space-charge forces. Experimental results confirm that high-intensity THz pulses with continuous spectral tunability from 7.8 to 30.8 THz can be achieved by simply adjusting the optical delay between the two beating lasers and the wiggler resonance.

## Results

The experiment was performed at the Shanghai Soft X-ray Free-Electron Laser facility (SXFEL) [34]. Fig. 1a illustrates the facility layout and experimental schematic. The linear accelerator (linac) consists of a photocathode injector, a laser heater (LH) system [35], a main accelerator, and two magnetic bunch compressors (BCs). A 400 $pC$ electron beam, initially generated by a photocathode gun with a full bunch length of 14.3 $ps$ and an energy of 4.6 MeV, was first accelerated to 115 MeV using the accelerator section A1 in the injector. The beam was then injected into a short undulator (the LH undulator, in Fig. 1a) where it interacted with a chirped frequency-beating laser pulse to acquire energy modulation.

A critical aspect of this setup is the preparation of the beating laser, which produces a tunable THz frequency signal through optical heterodyning of two linearly chirped, broadband laser pulses, as illustrated in Fig. 1a (see Methods for details). The energy modulation imposed on the electron beam replicates the intensity profile of the beating laser pulse, thereby imprinting a periodic structure on the beam's longitudinal phase space at the THz scale, as demonstrated by the simulation results in Fig. 1b (see Methods). Because both the beating laser and the photocathode drive laser originate from the same 800 nm Ti:Sapphire laser source, inherent synchronization is ensured. The frequency-beating method has been previously applied to conventional lasers [36] and storage rings [37]. The beating frequency is defined as [38]

$$f_0 = \frac{\mu\tau}{2\pi}, \qquad (1)$$

where $\mu$ is the chirp rate of the laser pulse. By adjusting the time delay $\tau$ between the two chirped lasers, $f_0$ can be continuously tuned without requiring additional modifications to the injector.

After modulation, the beam was further accelerated by A2 to approximately 235 MeV. A linear energy chirp was introduced by A2 and the downstream linearizer prior to the first bunch compressor (BC1), as shown in Fig. 1c. Following BC1, A3, and BC2, the electron beam was accelerated to 650 MeV and the bunch length was compressed by a total compression factor of approximately $C \approx 10$. Consequently, the THz energy modulation was converted into a periodic density modulation (i.e., microbunching, as shown in Fig. 1d) with a frequency

$$f = Cf_0 = \frac{C\mu\tau}{2\pi}. \qquad (2)$$

Sequentially, the electron beam was accelerated to approximately 1 GeV, with the energy chirp fully compensated by A4 (Fig. 1d). The longitudinal phase space of the electron beam was measured at the linac exit using an X-band deflecting cavity (TDX in Fig. 1a) combined with an energy spectrometer (Dipole in Fig. 1a). The measurement results are shown in Fig. 2. Notably, during this acceleration and compression processes, the THz modulation structure in the electron beam was not only preserved but also continuously enhanced by collective effects in the linac [39, 40] (see Methods).

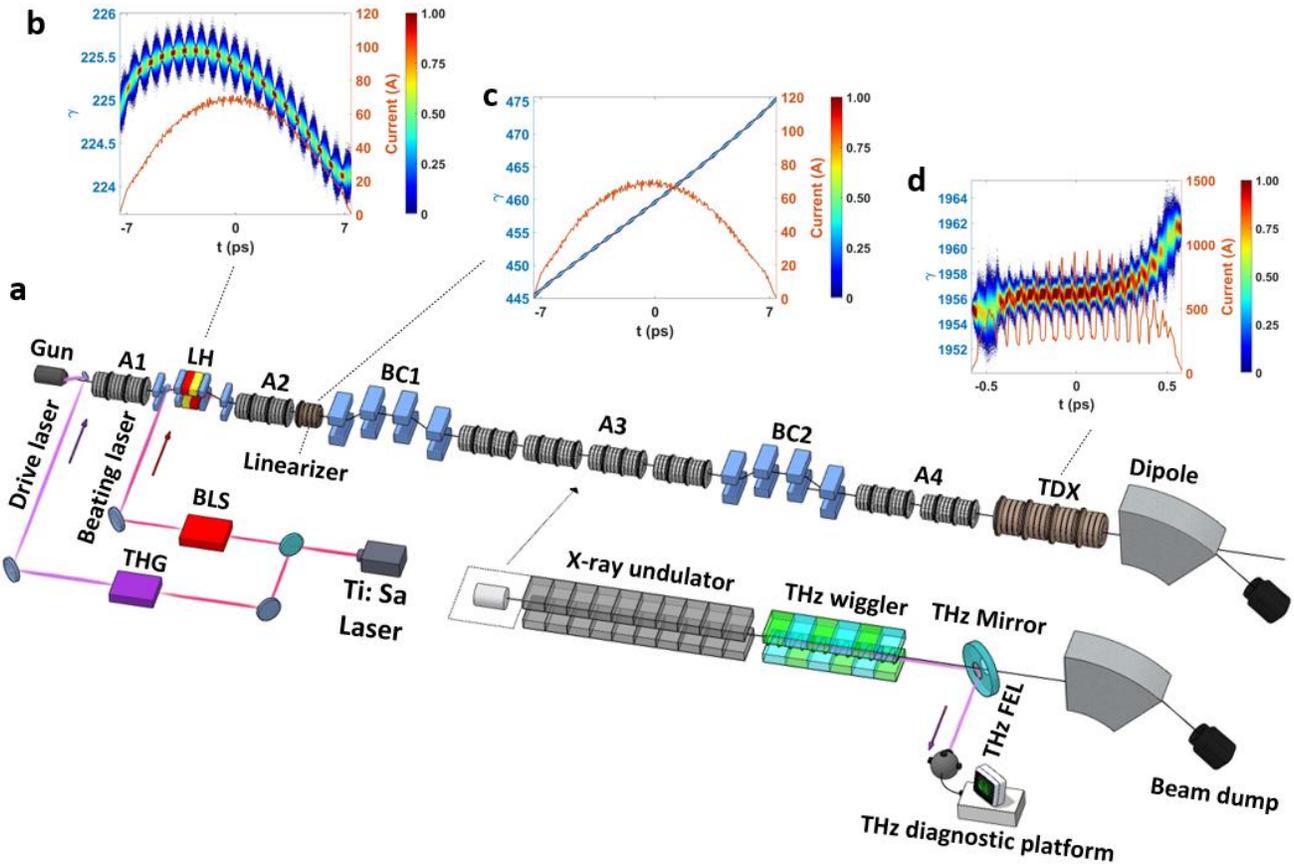

**Fig. 1: SXFEL facility layout and experimental schematic.** (**a**) The experimental beamline comprises a laser system, a linear accelerator (linac), a THz wiggler, and a THz diagnostic platform. An 800 nm pulse from the Ti:Sapphire laser is split into two branches: one is directed through a third-harmonic generation (THG) system to produce a 266 nm drive laser, while the other is routed to the beating laser system (BLS) to generate the required beating laser pulse. Inset panels illustrate simulation results detailing the evolution of the electron beam's longitudinal phase space. The orange lines indicate the projected density profiles. The electron beam first interacts with the beating laser in the laser heater (LH) to acquire an energy modulation (**b**); subsequently, a linear energy chirp is induced by accelerator section A2 and the linearizer (**c**). Following compression, the energy modulation is converted into a density modulation by the subsequent bunch compressors (BCs) (**d**); finally, the density-modulated beam is injected into the THz wiggler to generate narrowband THz radiation, which is then measured by the THz diagnostic platform.

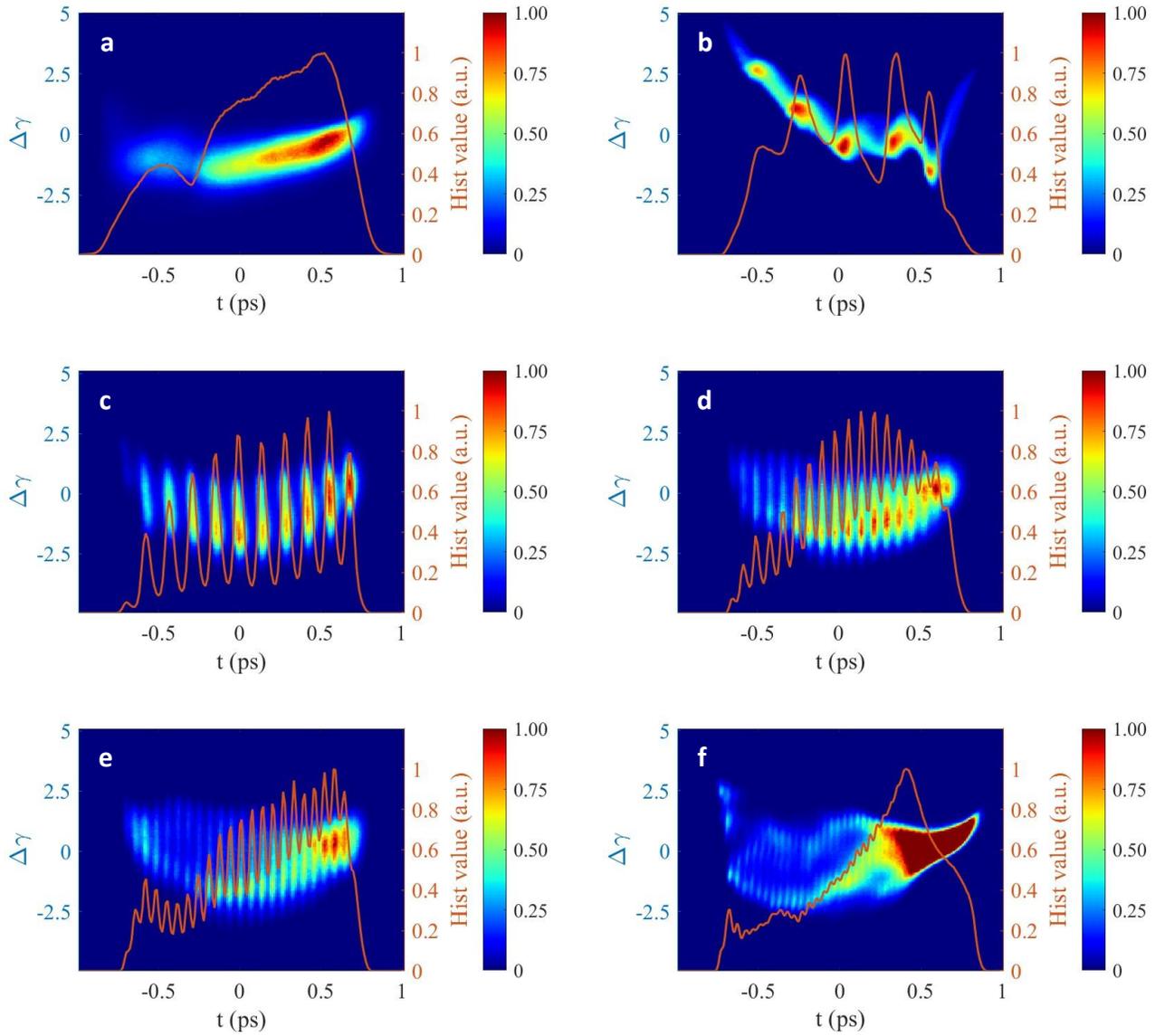

**Fig. 2: Longitudinal Phase Space Distributions and Projected Density Profiles of the Electron Beam. (a)** Measured phase space distribution with the beating laser turned off. Panels **(b–f)** display pre-bunched electron beams with bunching frequencies of 4 THz **(b)**, 10 THz **(c)**, 14.7 THz **(d)**, 19 THz **(e)**, and 24 THz **(f)**, achieved by tuning the time delay between the two beating lasers from 0.68 mm to 4.2 mm. The orange lines indicate the projected density profiles.

Fig. 2a displays the measured longitudinal phase spaces without the beating laser. Throughout the experiment, key parameters of the accelerator, such as beam energy and the compression factors of the two BCs, remained constant. By tuning on the beating laser and varying only the time delay $\tau$ from 0.68 mm to 4.2 mm, the bunching frequency was continuously tuned from approximately 4 THz to over 24 THz, as depicted in Figs. 2b-2f. Owing to the limited time resolution (~10 fs) of the longitudinal phase space measurements, the high-frequency bunching structures appear less distinct than the low-frequency ones. For bunching frequencies above 24 THz, the microbunching structure becomes increasingly difficult to resolve.

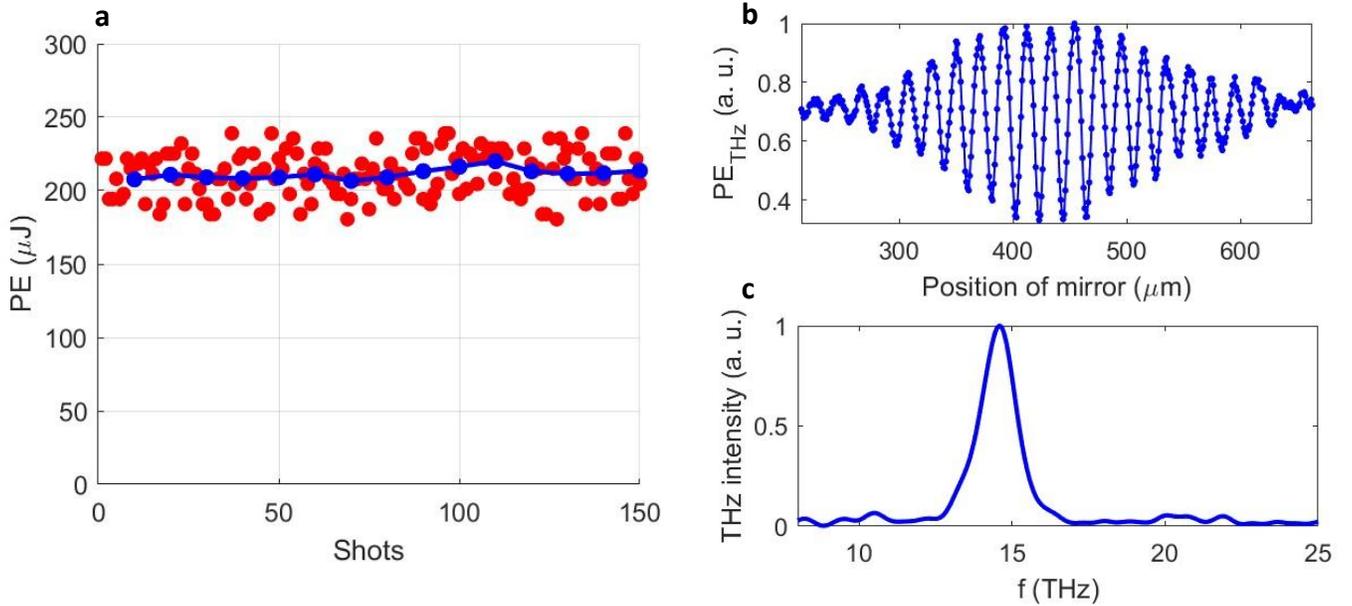

**Fig. 3: THz Radiation Performance at 14.7 THz.** **(a)** Individual pulse energies (red dots) and the corresponding average (blue line) measured over 150 consecutive shots. During measurements, a THz attenuator with a 30% transmission rate and a 15 THz band-pass filter were placed upstream of the detector. Autocorrelation function of the electric field obtained using a Michelson interferometer **(b)** and the corresponding Fourier transform **(c)**. The THz spectrum exhibits a bandwidth of approximately 8.4%.

After traversing an undulator section designed for X-ray FEL operation (which was deactivated during the experiment), the electron bunch trains were directed into a 5-meter-long electromagnetic wiggler (THz wiggler in Fig. 1a) with a period length of 0.28 m to generate narrow-band THz radiation. The radiation properties were characterized using a dedicated THz diagnostic platform (see Methods). Pulse energies were measured using a calibrated Golay cell detector, and the electric field distribution was obtained via a Michelson interferometer-based autocorrelation technique. As shown in Fig. 3a, THz pulse energies at different frequencies were statistically analyzed over 150 consecutive shots, revealing a maximum pulse energy of 239 $\mu J$ and a mean value of 211 $\mu J$, corresponding to a root mean square (RMS) relative fluctuation of 7.3%. Fig. 3b presents representative autocorrelation results, and Fig. 3c displays the corresponding radiation spectrum, centered at 14.7 THz with a full width at half-maximum (FWHM) bandwidth of 8.4%.

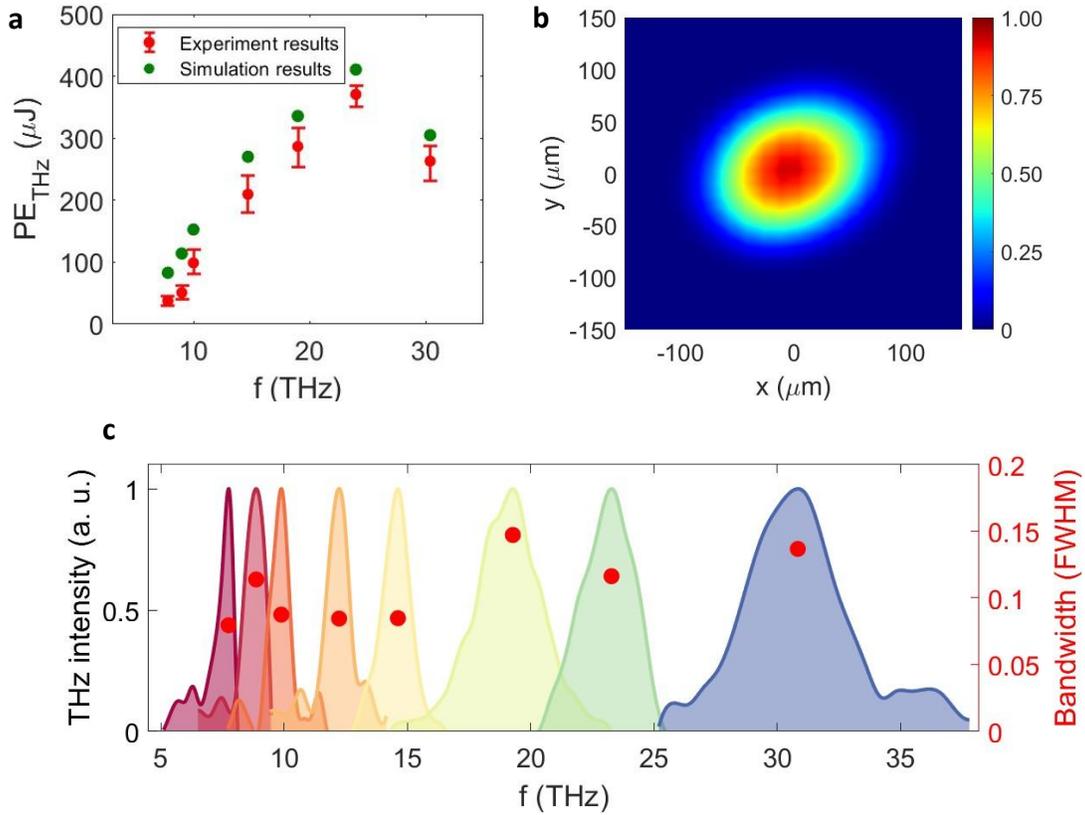

**Fig. 4: THz Radiation Measurements from 7.8 to 30.8 THz. (a)** Experimental and simulation results for the radiation pulse energies, averaged over 150 consecutive shots per frequency point; error bars represent the RMS intensity fluctuations. **(b)** The measured radiation spot at 19 THz at the focal point. **(c)** Spectra obtained from the Fourier transform of the autocorrelation functions, with the corresponding spectral bandwidths indicated by red dots.

The THz source frequency can be tuned by adjusting the time delay of the frequency-beating laser and the resonance of the THz wiggler. In our experimental setup, the frequency coverage was primarily limited by the magnetic field strength of the THz wiggler. With the beam energy fixed at 1 GeV, systematic variations in the laser delay from 0.68 mm to 4.20 mm produced continuous THz radiation spanning 10-24 THz. By altering the beam energy to, for example, 0.86 GeV or 1.2 GeV, the frequency range can be further extended from 4 THz to over 30 THz.

Figure 4 summarizes both measurement and simulation results across various THz frequencies. To isolate the narrowband THz radiation corresponding exclusively to the microbunching frequency, narrow-bandpass filters were placed before the Golay cell detector. Fig. 4a presents pulse energy measurements averaged over 150 consecutive shots per frequency point. The parameters of the frequency-beating laser and the BC compression factors were optimized for the 24 THz operation point, yielding a maximum pulse energy of approximately 385 $\mu J$. For other frequencies, only the laser delay and the THz wiggler resonance were adjusted, while all other parameters remained constant. Fig. 4b displays the typical THz radiation spot at 19 THz (focused by a THz focusing lens) measured by a THz camera. Fig. 4c shows the corresponding spectra at various frequencies. Spectral analysis reveals consistent relative FWHM bandwidths ranging from 7.7% to 14.7%, indicating well-preserved spectral coherence.

## Discussion

As shown in Fig. 1a, the THz wiggler was positioned downstream of the X-ray FEL undulator, functioning as a THz afterburner. This arrangement maximizes electron beam utilization and establishes a basis for future THz-pump X-ray-probe experiments [41]. However, operating with a 1 GeV high-energy electron beam optimized for XFEL pulse generation is suboptimal for this THz generation method. The two-stage compression in the linac complicates THz microbunching optimization, and any adjustments to BCs can adversely affect the performance of the linac. Consequently, our frequency tuning experiments were exclusively optimized for the 24 THz working point. This limitation resulted in suboptimal bunching for other frequencies (manifesting as either over-bunching or de-bunching), which in turn degraded the pulse energy relative to ideal values. In contrast, if THz-pump X-ray-probe capability is not essential, this method is better suited to a low-energy compact accelerator equipped with a single, small bunch compressor. In such a configuration, the shorter wiggler period permits a greater number of periods within the same wiggler length, thereby further enhancing the THz radiation power. FEL simulations suggest that this optimized beamline could extend the tuning range from 0.1 to 60 THz and boost pulse energies up to 6 $mJ$ with a 4 nC beam charge [38].

In summary, we have experimentally demonstrated a novel FEL-based approach for generating high-power, narrowband THz radiation with continuous frequency tunability. The experimental results reveal THz emissions spanning 7.8-30.8 THz, with maximal pulse energies reaching 385 $\mu J$ and spectral bandwidths (FWHM) maintained between 7.7% and 14.7%, achieved through amplification of the microbunching fundamental harmonics. The THz source frequency is readily tuned by adjusting the time delay of the frequency-beating laser and the THz wiggler resonance. Moreover, this method imposes no inherent limitations on repetition rate or beam charge, enabling the delivery of THz radiation with exceptional pulse energy and average power. The experimental results also demonstrate remarkable shot-to-shot stability in both pulse intensity and central frequency, offering new opportunities for both fundamental research and practical applications.

## Methods

**Frequency-beating Laser System**

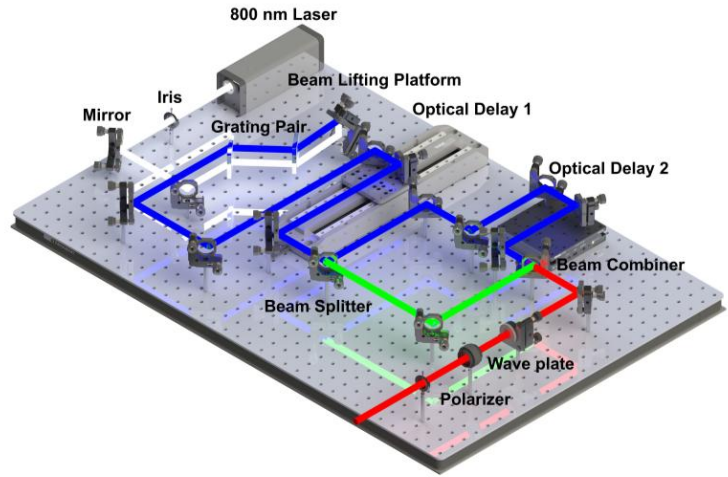

**Extended Data Fig. 1: The layout of frequency-beating laser system.**

Extended Data Fig. 1 illustrates the layout of the frequency-beating laser system. An 800 nm laser with an FWHM pulse duration of 100 fs is initially stretched to 22 $ps$ using a grating pair with 1500 mm$^{-1}$ line densities and a 100 mm vertical separation. The stretched beam is then elevated via a beam lifting platform and subsequently reflected off the same grating pair to cancel transverse dispersion. Next, the beam passes through a tunable optical delay line (Delay 1) to compensate for the temporal offset between the drive laser and the frequency-beating laser. The beam is then split into two branches by a beam splitter; one branch is further delayed using a second tunable optical delay line (Delay 2) to control the beating frequency. Finally, the two branches are recombined and directed into the LH undulator, producing a recombined pulse with a periodic field profile at the beating frequency ($f_0$). This system permits continuous tuning of $f_0$ from 0.1 to 4 THz by adjusting the optical delay from 0.22 to 9 mm.

**SXFEL Experimental Setup**

As depicted in Fig. 1a, the experimental setup begins with a 400 $pC$ electron beam generated from a photocathode gun, exhibiting a full bunch length ($L_{b0}$) of approximately 14.3 ps. The beam is first accelerated to 115 MeV using accelerator section A1 (S-band RF structures) before being injected into the LH undulator, which has a 50 mm magnetic period and a total length of 0.5 m. Concurrently, the frequency-beating laser imprints an energy modulation onto the beam, replicating its periodic intensity profile at the THz scale.

Subsequently, the beam is further accelerated by section A2, comprising two S-band and one X-band accelerating sections, to approximately 235 MeV. An energy chirp is introduced via off-crest S-band operation and then linearized with an off-crest X-band section, establishing an initial chirp rate $h_0$. The chirped beam is then compressed by a magnetic chicane (BC1 in Fig. 1a) whose dispersion ($R_{56}^1$) is adjustable between 0 and 50 mm. The compression factor, defined as $C_1 = \frac{L_{b1}}{L_{b0}} = \frac{1}{|1+h_0 R_{56}^1|}$, compresses the bunch length to approximately 2-3 ps and adjusts the chirp to $h_1 = C_1 h_0$. As a result, the initial THz energy modulation is converted into a periodic density modulation with frequency $f_1 = C_1 f_0$. This density-modulated beam is further accelerated to 650 MeV through four C-band accelerating sections (A3). During this phase, collective effects, dominated by longitudinal space charge (LSC), enhance the THz modulation by converting the density modulation into an additional energy modulation of amplitude $\Delta\gamma(k_1)$, where $k_1 = \frac{c}{f_1}$.

The beam then passes through a second magnetic chicane (BC2 in Fig. 1a) with compression factor $C_2 = \frac{L_b}{L_{b1}} = \frac{1}{|1+h_1 R_{56}^2|}$, where $L_b$ is the bunch length after BC2 and $R_{56}^2$ is the dispersion strength. This stage further

converts the THz energy modulation into a density modulation with frequency $f = C_2 f_1$ (or equivalently, $k = k_1/C_2$) and results in a bunching factor $b_f(k_1/C_2)$. The overall gain of the bunching process is given by [35, 39, 40]

$$G(k) = \frac{b_f(k)}{b(k_1)} = \left| C_2 k_1 R_{56}^2 \frac{\Delta\gamma(k_1)}{\gamma b(k_1)} \right| e^{-[(C_2 k_1 R_{56}^2 \sigma_E)^2/2E^2]},$$

where $\gamma = E/mc^2$ and $\sigma_E$ expresses the uncorrelated energy spread. Following this, the remaining sections of the linac (A4 with C-band structures) accelerate the beam to approximately 1 GeV while compensating for the energy chirp. Finally, the longitudinal phase space is characterized at the linac exit using an X-band transverse deflecting cavity in combination with an energy spectrometer.

The density-modulated beam is then directed into a 5.04 m electromagnetic THz undulator with a 0.28 m magnetic period. The SXFEL facility comprises two undulator lines (SBP and SUD), with the THz undulator installed along the straight SBP beamline. During the experiments, the gap of the X-ray undulator (as seen in Fig. 1a) is maintained at its maximum opening.

**THz Diagnostic Platform**

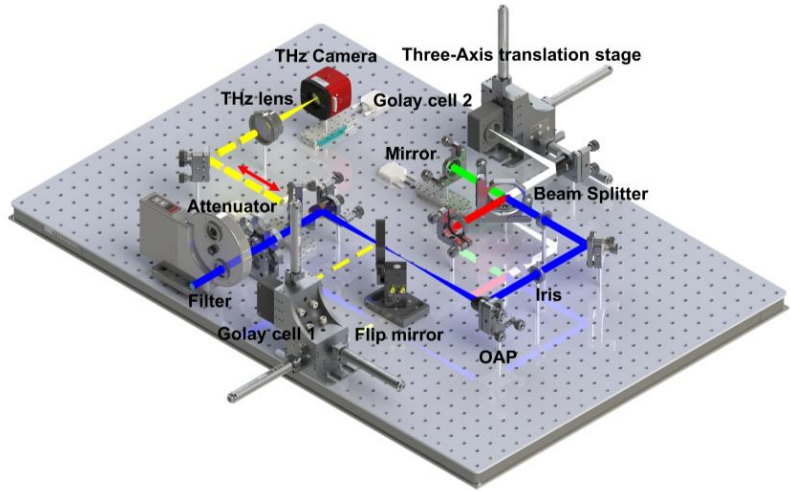

**Extended Data Fig. 2: The layout of THz diagnostic platform.**

Extended Data Fig. 2 presents the schematic of the THz diagnostic platform. In this setup, the THz radiation, co-propagating with the electron beam, is first directed onto a gold-coated THz deflector mirror with a 2 mm circular aperture. While the electron beam passes directly through this aperture, most of the THz radiation is reflected into a diamond vacuum window and enters the diagnostic platform. At the platform entrance, a band-pass filter (BPF) or low-pass filter (LPF) and switchable THz attenuators (with transmission rates of 30%, 10%, and 1%) are installed. For pulse energy measurements, an LPF with an 18.9 THz cutoff is employed for the 7.8 THz and 24 THz signals (by subtracting the pulse energy measured with the LPF), while BPFs with central frequencies of 10, 15, 19, and 30 THz are used for other frequencies. The THz radiation is subsequently either collected by a gold-coated off-axis parabolic mirror (OAP) or focused on a THz camera (Dolphin Optics: Beam 2000) via a movable mirror and a THz focusing lens (Swiss THz optics F0.7) for beam spot measurements. In the OAP path, a flip mirror directs the THz radiation either to the pulse energy measurement path, where a Golay cell (TYDEX, model GC-1D) mounted on a three-dimensional displacement stage is placed at the focal point, or to the power density spectrum path. In the latter configuration, the radiation is collimated by a secondary OAP and then introduced into a Michelson interferometer, where a second Golay cell captures the interference signal. Fourier transformation of this signal yields the THz power density spectrum.

**Golay Cell Calibration**

Two Golay cells (TYDEX, model GC-1D) were calibrated by using a continuous wave $CO_2$ laser with a wavelength of 10.6 $\mu m$ at the Laser Electron Gamma Source (SLEGS) beamline of the Shanghai Synchrotron Radiation Facility (SSRF) [42]. As shown in Extended Data Fig. 3a, the laser beam is split into two paths after beam focusing by ZnSe lens: one path directs the beam to a calibrated infrared dynamometer for pulse energy measurement, after which the dynamometer is replaced by the Golay cell; the second path is monitored by an infrared detector to record the pulse shape in real time. By varying the repetition rate and pulse duration, the responses of Golay cell-1 and Golay cell-2 were determined to be 1.83 $\mu J/V$ and 3.29 $\mu J/m$, respectively (see Extended Data Figs. 3b and 3c).

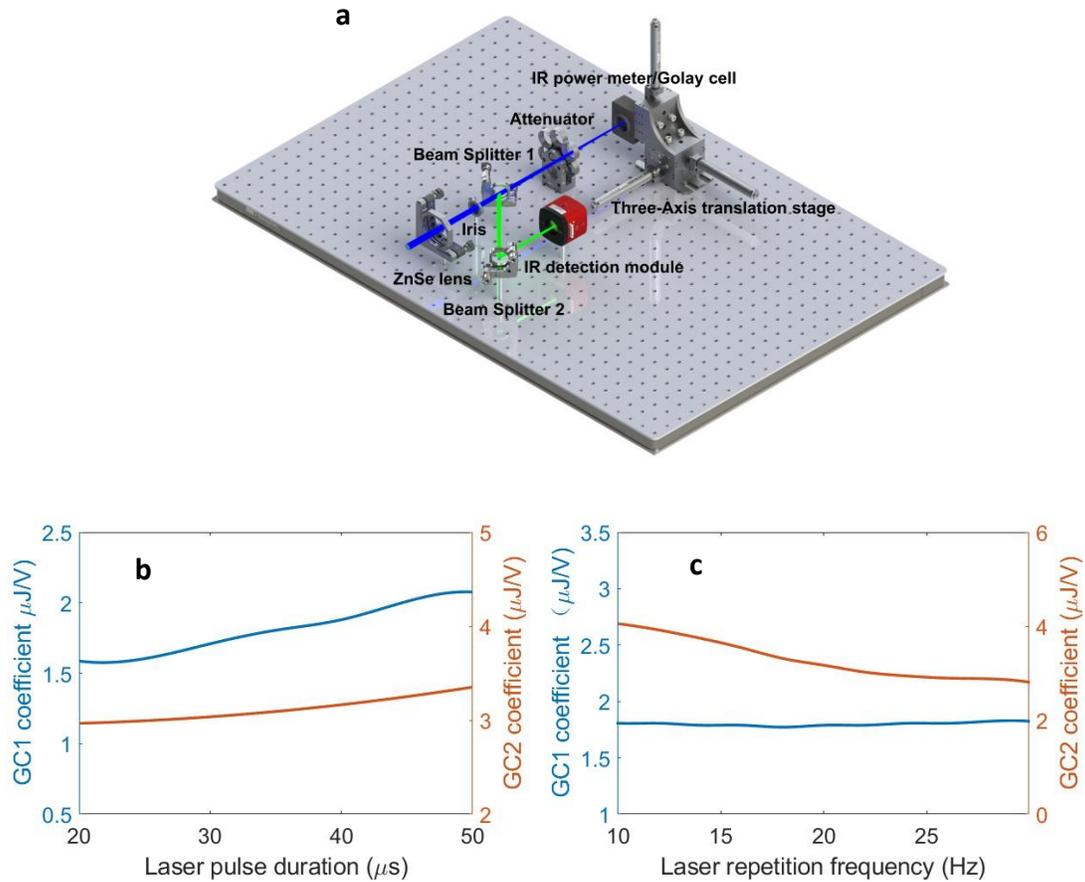

**Extended Data Fig. 3:** The layout of Golay cells calibration optical path (**a**), the calibrated results of Golay cell-1 (**b**) and Golay cell-2 (**c**) by changing the pulse duration and repetition rate.

**Simulation Method**

Start-to-end simulations were performed for the experiments. Beam dynamics from the photocathode to the linac exit were simulated using the particle-tracking codes ASTRA [43] and ELEGANT [44], which incorporate longitudinal space charge and coherent synchrotron radiation effects. The generation of THz FEL radiation was simulated using the time-dependent code GENESIS [45].

## Data Availability

The data that support the plots within this paper and other findings of this study are available from the corresponding author upon reasonable request.

## Acknowledgements

The authors thank Min Chen and Huachun Zhu for helpful discussions. This work was supported by the National Natural Science Foundation of China (NSFC grant no. 12275340, 12105347, 12435011), CAS Project for Young Scientists in Basic Research (YSBR-115), Shanghai Municipal science and Technology Major Project and Innovation Program of Shanghai Advanced Research Institute, CAS (2024CP001).

## Contributions

C. F, K. Z and Y. K conceived and designed the experiments. K. Z and Y. K conducted the experiments with the help from Z. W, T. L, Y. W, Z. G, H. L, X. W, T. L, Z. Q and with the software and hardware supports by C. Y, W. Y, H. X, J. C, T. L, X. L, J. W, H. Z, F. G, L. S, Y. Z, Y. W, C. X, X. W, J. X, B. L. The THz diagnostic system is designed and constructed by T. L, L. Y, Q. T, Y. W and Y. K. The simulations on beam dynamics were performed by Y. K. and Z. W. The manuscript was written by C. F., K. Z. and Y. K. with contributions from Z. W. Management and oversight of the project was provided by C. F., B. L. and Z. Z.

## Ethics declarations

Competing interests

The authors declare no competing interests.